\documentstyle[12pt,qqaalart]{article}

\author{Jos\'e Antonio Belinch\'on\thanks{E-mail: abelinchon@caminos.recol.es}\\
Grupo Interuniversitario de An\'alisis Dimensional.\\
Dept. de F\'isica ETS Arquitectura UPM\\
Av. Juan de Herrera 4 Madrid 28040 Espa\~na}
\title{Cosmological models with bulk viscosity in presence of adiabatic matter creation and with  $G$, $c$ and $\Lambda $ variables
}
\input tcilatex

\QQQ{Language}{
American English
}

\begin{document}

\maketitle
\begin{abstract}
Some properties of cosmological models with a time variable bulk viscous
coefficient in the presence of adiabatic matter creation and $G,$ $c,$ $%
\Lambda $ variables are investigated in the framework of flat FRW line
element. We trivially find a set of solutions through Dimensional Analysis.
In all the studied cases it is found that the the behaviour of these
''constants'' is inversely proportional to the cosmic time.
\end{abstract}

\section{\bf Introduction.}

Recently several models with FRW metric, where ''constants'' G and Lambda
are considered as dependent functions of time $t$ have been studied. For
these models, whose energy-momentum tensor describes a perfect fluid, it was
demonstrated that $G\propto t^\alpha $, where $\alpha $ represents a certain
positive constant that depends on the state equation imposed while $\Lambda
\propto t^{-2}$ is indepedent of the state equation (see \cite{A}). More
recently this type of model was generalized by Arbab (see \cite{AB}) who
considers a fluid with bulk viscosity (or second viscosity in the
nomenclature of Landau (see \cite{L})). The role played by the viscosity and
the consequent dissipative mechanism in cosmology has been studied by many
authors. The heat represented by the great entropy per baryon in the cosmic
background radiation provides a good indication of the early universe and a
possible explanation of the huge entropy per baryon which is believed to
have been generated by physical dissipative processes acting at the
beginning of evolution. These dissipative processes may be responsible for
the smoothing out of initial anisotropies.

In the models studied by Arbab constants $G$ and $\Lambda $ are substituted
by scalar functions that depend on time $t$. The state equation that governs
the bulk viscosity is: $\xi \propto \xi _0\rho ^\gamma $ where $\gamma $ is
a certain indeterminate constant for the time being $\gamma \in \left[
0,1\right] $. Amongst all the possible $\gamma $ we are only going to
concentrate on $\gamma $$=1/2$ since, as pointed out by Golda et al. (see 
\cite{Z}) upon demonstrating that for an adequate election of the state
equation the viscous models are topologically equivalent (structurally
stable) to the classic FRW. Golda et al show that the bulk viscous fluids
with FRW symmetries can structurally approximate the dynamics of the classic
FRW. They found that approximation only takes place when the parameter of
the bulk viscosity follows the state equation $\xi \propto \xi _0\rho ^{1/2}$%
. They seek spatially homogeneous and isotropic solutions with bulk
viscosity that could be approximated in a structurally stable way to the
dynamics that describe the classic FRW described by an ordinary perfect
fluid. More specifically, they looked for such spatially homogeneous and
isotropic solutions which after being disturbed by the dissipative parameter
have a dynamics that is topologically equivalent to the classic FRW models.
Therefore the only viscous models that can be approximated to the classic
FRW in a structurally stable way are those that follow a state equation $\xi
\propto \xi _0\rho ^{1/2}$ because of their viscous parameter.

The first authors that have studied the convenience of also considering the
variation of ''constant'' $c$, the speed of light, as a scalar function
depending on time $t$, to solve some of the problems that classic FRW models
present, have been (see \cite{TR},\cite{P}). Others (see \cite{BA} and \cite
{AL}) demonstrate through a variational formulation that the resulting field
equations continue being the Friedmann ones if a FRW metric is considered.
In these papers it is not so much the accurate calculation of such a
variation for c that is taken into consideration as the examination of the
benefits of a modification in the field equations. Thus, these models are
able to explain the problem of the horizon and the flatness of the universe,
in the same way as the inflationary models.

Following Arbab's line, the author of this article (see \cite{T1}) attempts
to generalize the models with $G,c$ and $\Lambda $ variables through the
consideration of a flow with bulk viscosity. In this type of models, (the
structurally stable $\gamma =1/2$) inspite of considering viscous fluids,
that involve a natural production of entropy (irreversible processes), it
demonstrates that the entropy continued being constant (as in the classic
FRW). In this paper we consider both mechanisms of matter production and
mechanisms of entropy. To attempt to solve this problem we consider in this
paper so much mechanisms of matter production as of entropy. The problem of
matter creation in the universe has been studied by many authors following
Prigogine and co-workers pioneering ideas (see \cite{PRI}). We will follow
Lima et al's work. (see (\cite{LI}). This problem in the bulk viscous fluids
framework has been studied by Desikan (\cite{DE}). The general idea is the
following: to consider the universe as an open system, the creation of
matter itself generatersemtropy and thus the second law of the
thermodynamics can be incorporated into the field of equations. In the case
of open systems the number of particles in a given volume is not fixed and
though we consider that the transformation is adiabatic, it is demonstrated
that the entropy is growing when the matter creation acts as an internal
energy mechanism. In this paper we calculate the variation of ''constants'' $%
G,c$ and $\Lambda $ in the framework of a model with FRW symmetries, with $%
k=0$ i.e. flat, with adiabatic matter creation and bulk viscosity as
separate irreversible processes. We consider the processes of adiabatic
matter creation (see \cite{LI}) to attempt to solve the problem of entropy.
We wish to draw attention to how the use of the Dimensional Analysis (D.A.)
(see \cite{B}) enables us to find the solution to our model in a trivial
way. We do not want to carry out a meticulous study of the cosmological
implications of the results obtained for each of the parameters on which the
solution depends ($\omega ,\beta ,\gamma $) (postponing this paragraph for a
subsequent article) but simply to demonstrate them. We will show that for an
adequate election of such parameters we obtain the same behaviour for the
principal quantities as those obtained for the classic FRW models. The
behaviour obtained for the ''constants'' is that for these parameters they
always vary in a way inversely proportional to the time. These results are
compared to those of other authors.

The paper is organised as follows: In the second section the governing
equations of our model will be shown and considerations on the followed
dimensional method will be made. In the third section we will make use of
the D.A. (Pi theorem) to obtain a solution to the principal magnitude that
appears in the model and finally in the fourth section we will present two
particular cases of the obtained solutions together with some conclusions.

\section{\bf The model.}

For a flat universe $k=0$ with FRW symmetries i.e. we assumed homogeneity
and isotropy and therefore there will not be spatial variations of the
''constants'' $G,c$ and $\Lambda $ solely temporary. We suppose equally that
our fluid is bulk viscous (second viscosity) and we consider mechanisms of
creation of matter. With these suppositions the equations that govern the
model are as follows: 
\begin{equation}
\label{e1}-2\frac{f\,^{\prime \prime }}{f\,}-\frac{(f\,^{\prime })^2}{f\,^2}%
= \frac{8\pi G(t)}{c^2(t)}(p+p_c)+c^2(t)\Lambda (t)\ \ 
\end{equation}

\begin{equation}
\label{e2}\frac{(f\,^{\prime })^2}{f\,^2}=\frac{8\pi G(t)}{3\,c^2(t)}\rho
+c^2(t)\Lambda (t)\qquad \quad \ 
\end{equation}
\begin{equation}
\label{e3}n^{\prime }+3nH-\psi =0 
\end{equation}
where $n$ measures the particles number density, $\psi $ is the function
that measures the matter creation, $H=f^{\prime }/f$ represents the Hubble
parameter ($f$ is the scale factor that appears in the metrics), $p$ is the
thermostatic pressure, $\rho $ is energy density and $p_c$ is the pressure
that generates the matter creation.

The creation pressure $p_c$ depends on the function $\psi $. For adiabatic
matter creation this pressure takes the following form (\cite{LI}): 
\begin{equation}
\label{w2}p_c=-\left[ \frac{\rho +p}{3nH}\psi \right] 
\end{equation}
The state equation that we will use is the known expression 
\begin{equation}
\label{w3}p=\omega \rho 
\end{equation}
where $\omega =const.$ $\omega \in \left[ 0,1\right] $ physically realistic
equations, making in this way that the energy-momentum tensor $T_{ij}$
verify the energy conditions.

We need to know the exact form of the function $\psi $ the one which is
determined from a more fundamental theory than involves quantum processes.
We assumed that this function continues the following law: 
\begin{equation}
\label{w5}\psi =3\beta nH 
\end{equation}
here we are following to Lima et al (see \cite{LI}) for other treatment \cite
{DE} while Prigogine et al \cite{PRI} follows this other law $\psi =\kappa
H^2)$ where $\beta $ is a dimensionless constant (if $\beta =0$ then there
is matter creation since $\psi =0)$ that it is given presumably by models of
particles physics of matter creation. Physically one hope that the most
interesting situations emerge during phases in those which i.e. $\beta
\approx 1$ will be of the order of unity.

The conservation principle carries us to the following expression: 
\begin{equation}
\label{w4}\rho ^{\prime }+3(\omega +1)\rho \frac{f^{\prime }}f=(\omega
+1)\rho \frac \psi n 
\end{equation}

Integrating the equation (\ref{w4}) we obtain the following relationship
between energy density and the radius of the universe and what is more
important the constant of integration necessary for our subsequent
calculations: 
\begin{equation}
\label{e4}\rho =A_{\omega ,\beta }f^{-3(\omega +1)(1-\beta )} 
\end{equation}
where $A_{\omega ,\beta }$ is the constant of integration that depends on
the state equation that is considered i.e. of the constant $\omega $ and of
the constant $\beta $ that measures the matter creation.

The effect of the bulk viscosity in the equations is shown replacing $p$ by $%
p-3\xi H$ where $\xi $ follow the law $\xi =\xi _0\rho ^\gamma $ (see \cite
{W},\cite{AB} and \cite{DE}). This last state equation, in our opinion, it
does not verify the homogeneity principle by this reason we modify it by: 
\begin{equation}
\label{n2}\xi =k_\gamma \rho ^\gamma 
\end{equation}
where the constant $k_\gamma $ causes that this equation yes will be
dimensionally homogeneous for any value of $\gamma .$

The dimensional analysis that we followed needs to make the following
distinctions, we need to know beforehand the set of fundamental quantities
together with that of unavoidable constants (in the nomenclature of
Barenblatt are designated as governing parameters). In this case the single
one fundamental quantity that appears in the model is the cosmic time $t$ as
can be deduced with facility of the homogeneity and isotropy supposed for
the model. The unavoidable constants of the model are the constant of
integration $A_{\omega ,\beta }$ that depends on the state equation $\omega $
and of the mechanisms of matter creation $\beta $ and the constant $k_\gamma 
$ that controls the influence of the viscosity in the model.

In a previous work (\cite{T}) was calculated the dimensional base of this
type of models, being this $B=\left\{ L,M,T,\theta \right\} $ where $\theta $
represents the dimension of the temperature. The dimensional equation of
each one of the governing parameters are:%
$$
\begin{array}{c}
\left[ t\right] =T\quad \left[ A_{\omega ,\beta }\right] =L^{3(\omega
+1)(1-\beta )-1}MT^{-2}\qquad \\ 
\left[ k_\gamma \right] =L^{\gamma -1}M^{1-\gamma }T^{2\gamma -1} 
\end{array}
$$
All the derived quantities or governed parameters in the nomenclature of
Batrenblatt we will calculate them in function of these quantities (the
governing parameters), that is to say, in function of the cosmic time $t$
and of the two unavoidable constants $k_\gamma $ and $A_{\omega ,\beta }$
with respect to the dimensional base $B=\left\{ L,M,T,\theta \right\} .$

\section{{\bf {Solutions through D.A.}}}

We go to calculate through dimensional analysis D.A. i.e. applying the Pi
Theorem, the variation of $G(t)$ in function of $t$, the speed of light $%
c(t),$ the energy density $\rho (t),$ the radius of the universe $f(t),$ the
temperature $\theta (t)$, the entropy $S(t)$ and the entropy density $s(t),$%
the viscosity coefficient $\xi $$(t),$ the particle number density $%
n(t)\propto f^{-3}$ and finally the variation of the cosmological
''constant'' $\Lambda (t).$

The dimensional method carries us to (see \cite{B})

\subsection{{\bf Calculation of }$G(t)$}

As we have indicated above, we are going to accomplish the calculation of
the variation of $G$ applying the Pi theorem. The quantities that we
consider are: $G=G(t,k_\gamma ,A_{\omega ,\beta }).$ with respect to the
dimensional base $B=\left\{ L,M,T,\theta \right\} .$ We know that $\left[
G\right] =L^3M^{-1}T^{-2}$%
$$
\left( 
\begin{array}{ccccc}
& G & t & k_n & A_\omega \\ 
L & 3 & 0 & \gamma -1 & 3(\omega +1)(1-\beta )-1 \\ 
M & -1 & 0 & 1-\gamma & 1 \\ 
T & -2 & 1 & 2\gamma -1 & -2 
\end{array}
\right) 
$$
we obtain a single monomial that leads to the following expression for $G$%
\begin{equation}
\label{r1} 
\begin{array}{c}
G\propto A_{\omega ,\beta }^{\frac 2{3(\omega +1)(1-\beta )}}k_n^{
\frac{2+3(\omega +1)(1-\beta )}{3(\omega +1)(1-\beta )(\gamma -1)}} \\ 
t^{-4-\left[ \frac{2+3(\omega +1)(1-\beta )}{3(\omega +1)(1-\beta )(\gamma
-1)}\right] } 
\end{array}
\end{equation}

\subsection{\bf Calculation of $c(t)$}

$c(t)=c(t,k_\gamma ,A_{\omega ,\beta })$ where $\left[ c\right]
=LT^{-1}\Longrightarrow $

{\bf 
\begin{equation}
\label{r2} 
\begin{array}{c}
c(t)\propto A_{\omega ,\beta }^{\frac 1{3(\omega +1)(1-\beta )}}k_n^{\frac
1{3(\omega +1)(1-\beta )(\gamma -1)}} \\ 
t^{-1-\left[ \frac 1{3(\omega +1)(1-\beta )(\gamma -1)}\right] } 
\end{array}
\end{equation}
}

\subsection{\bf Calculation of energy density $\rho (t)$}

$\rho =\rho (t,k_\gamma ,A_{\omega ,\beta })$ with respect to the
dimensional base $B,$ where $\left[ \rho \right] =L^{-1}MT^{-2}$%
\begin{equation}
\label{r3}\rho \propto k_n^{\frac 1{1-\gamma }}t^{\frac 1{\gamma -1}} 
\end{equation}
we observe that this relationship shows us that energy density does not
depend neither on the state equation $\omega $ nor on the mechanisms on
creation of matter i.e. does not depend on the constant $A_{\omega ,\beta }$
solely on the viscosity of the fluid.

\subsection{\bf Calculation of the radius of the universe $f(t).$}

$f=f(t,k_\gamma ,A_{\omega ,\beta })$ where $\left[ f\right]
=L\Longrightarrow $%
\begin{equation}
\label{r4}f\propto A_{\omega ,\beta }^{\frac 1{3(\omega +1)(1-\beta
)}}k_n^{\frac 1{3(\omega +1)(1-\beta )(\gamma -1)}}t^{\frac{-1}{3(\omega
+1)(1-\beta )(\gamma -1)}} 
\end{equation}
We can observe that:%
$$
\begin{array}{c}
q=- 
\frac{f^{\prime \prime }f}{\left( f^{\prime }\right) ^2}= \\ -\left[ 3\beta
(\omega +1)(\gamma -1)-3n(\omega +1)+3\omega +4\right] 
\end{array}
$$
$$
H=\frac{f^{\prime }}f=-\left( \frac 1{3(\omega +1)(1-\beta )(\gamma
-1)}\right) \frac 1t 
$$

\subsection{\bf Calculation of the temperature $\theta (t).$}

$\theta =\theta (t,k_\gamma ,A_{\omega ,\beta ,}k_B)$ where $k_{B\text{ }}$
is the Bolztmann constant : $\left[ \theta \right] =\theta $ and $\left[
k_B\theta \right] =L^2MT^{-2}\Longrightarrow $%
\begin{equation}
\label{r5}k_B\theta \propto A_{\omega ,\beta }^{\frac{-1}{(\omega
+1)(1-\beta )}}k_n^{\frac{1-(\omega +1)(1-\beta )}{(\omega +1)(1-\beta
)(\gamma -1)}}t^{-\left[ \frac{1-(\omega +1)(1-\beta )}{(\omega +1)(1-\beta
)(\gamma -1)}\right] } 
\end{equation}

\subsection{{\bf Calculation of the entropy }$S(t):$}

$S=s(t,k_\gamma ,A_{\omega ,\beta },a)$ where $a$ is the radiation constant. 
$\left[ S\right] =L^2MT^{-2}\theta ^{-1}$%
\begin{equation}
\label{r6}S\propto A_{\omega ,\beta }^{\frac{-1}{(\omega +1)(1-\beta )}}k_n^{
\frac{1-\frac 34(\omega +1)(1-\beta )}{(\omega +1)(1-\beta )(\gamma -1)}%
}t^{-\left[ \frac{1-\frac 34(\omega +1)(1-\beta )}{(\omega +1)(1-\beta
)(\gamma -1)}\right] }a^{\frac 14} 
\end{equation}

\subsection{{\bf Calculation of the entropy density }$s(t):$}

$s=s(t,k_\gamma ,A_{\omega ,\beta },a)$ where $a$ is the radiation constant. 
$\left[ s\right] =L^{-1}MT^{-2}\theta ^{-1}$%
\begin{equation}
\label{r7}s\propto A_{\omega ,\beta }^0k_n^{\frac 3{4(\gamma -1)}}t^{-\left[
\frac 3{4(\gamma -1)}\right] }a^{\frac 14} 
\end{equation}

\subsection{\bf Calculation of the viscosity coefficient $\xi $$(t):$}

$\xi $$=\xi (t,k_\gamma ,A_{\omega ,\beta })$ where $\left[ \xi \right]
=L^{-1}MT^{-1}$%
\begin{equation}
\label{r8}\xi \propto k_n^{\frac 1{1-\gamma }}t^{\frac{-\gamma }{\gamma -1}} 
\end{equation}

\subsection{\bf Calculation of the cosmological constant: $\Lambda (t).$}

$\Lambda =\Lambda (t,k_\gamma ,A_{\omega ,\beta })$ where $\left[ \Lambda
\right] =L^{-2}$%
\begin{equation}
\label{r9}\Lambda \propto A_{\omega ,\beta }^{\frac{-2}{3(\omega +1)(1-\beta
)}}k_n^{\frac{-2}{3(\omega +1)(1-\beta )(\gamma -1)}}t^{\frac 2{3(\omega
+1)(1-\beta )(\gamma -1)}} 
\end{equation}

\section{\bf Different cases.}

All the following cases can be calculated without difficulty. But as we have
indicate in the first section we are going to centre our attention only in
those models that follow the law $\xi =k_\gamma \rho ^{1/2}$ i.e. $\gamma
=(1/2)$ that corresponds to models that are topologically equivalent to the
classic FRW. We are going to study two models with $\gamma $$=(1/2),$ one
with $\omega =1/3$ that corresponds to a universe with radiation
predominance and other with $\omega =0$ corresponding to a universe with
matter predominance.

\subsection{$\gamma $$=1/2$ and $\omega =1/3$}

$$
\begin{array}{c}
G\propto A_\omega ^{\frac 1{2(1-\beta )}}k_n^{-2-\frac 1{(1-\beta
)}}t^{-2+\frac 1{(1-\beta )}}\quad \\ 
\qquad c\propto A_\omega ^{\frac 1{4(1-\beta )}}k_n^{
\frac{-1}{2(1-\beta )}}t^{-1+\frac 1{2(1-\beta )}}\quad \qquad \\ \Lambda
\propto A_\omega ^{-\frac 1{2(1-\beta )}}k_n^{\frac 1{(1-\beta )}}t^{-\frac
1{(1-\beta )}}\qquad \\ 
\text{if }\beta =0\Rightarrow G\propto t^{-1},c\propto t^{-1/2},\Lambda
\propto t^{-1} 
\end{array}
$$
$c\propto t^{-1/2}$ also has been obtained by Barrow \cite{BA} and Troiskii 
\cite{TR}, with respect to the rest of the quantities we have obtain the
same behaviour that Lima et al. \cite{LI}%
$$
\begin{array}{c}
\rho \propto k_n^2t^{-2}\quad \quad \quad \rho \propto t^{-2}\quad \\ 
k_B\theta \propto A_\omega ^{\frac 3{4(1-\beta )}}k_n^{2-\frac 3{2(1-\beta
)}}t^{-2+\frac 3{2(1-\beta )}}\qquad 
\end{array}
$$
$$
f\propto A_\omega ^{\frac 1{4(1-\beta )}}k_n^{-\frac 1{2(1-\beta )}}t^{\frac
1{2(1-\beta )}}\qquad \qquad 
$$
$$
a^{\frac{-1}4}S\propto A_\omega ^{\frac 3{4(1-\beta )}}k_n^{-\frac{3\beta }{%
2(1-\beta )}}t^{\frac{3\beta }{2(1-\beta )}}\qquad 
$$
$$
a^{\frac{-1}4}s\propto A_\omega ^0k_n^{\frac 32}t^{-\frac 32}\qquad s\propto
t^{^{-\frac 32}} 
$$
$$
\xi \propto k_n^2t^{-1}\quad \quad \quad \xi \propto t^{-1} 
$$
$$
\text{Si }\beta =0\Rightarrow \theta \propto t^{-1/2},f\propto
t^{1/2},S\propto t^0=const. 
$$
With respect to the thermodynamic behavior, the matter creation formulation
considered here is a clear consequence of the nonequilibrium thermodynamic
in presence of a gravitational field. We see that the $\beta $ parameter
works in the opposite sense to the expansion, that is, reducing the cooling
rate with respect to the case where there is no matter creation. A very
meaningful result is the fact that the spectrum of this radiation cannot be
distinguished from the usual blackbody spectrum at the present epoch (see 
\cite{LI}). Therefore models with adiabatic matter creation can be
compatible with the isotropy currently observed in the spectral distribution
of the background radiation. We observe equally, that the obtained model is
clearly irreversible (classic FRW is reversible). We want also to express
the fact that all the important thermodynamic quantities of the classic FRW
models are recovered if we made $\ \beta =0.$ (see \cite{T})%
$$
f\propto t^{1/2},\ \rho \propto t^{-2},\ \theta \propto t^{-1/2},\
S=const.,\ s\propto t^{-3/2} 
$$
Finally it is interesting to stick out that the presented model may
significantly alter the predictions that make the classic FRW on the
abundance of elements. Such result puts a possible limitation to the values
that could take the $\beta $ parameter.

\subsection{$\gamma $$=1/2$ and $\omega =0:$}

$$
\begin{array}{c}
G\propto A_\omega ^{\frac 2{3(1-\beta )}}k_n^{-2-\frac 4{3(1-\beta
)}}t^{-2+\frac 4{3(1-\beta )}}\quad \\ 
\qquad c\propto A_\omega ^{\frac 1{3(1-\beta )}}k_n^{
\frac{-2}{3(1-\beta )}}t^{-1+\frac 2{3(1-\beta )}}\quad \qquad \\ \Lambda
\propto A_\omega ^{-\frac 2{3(1-\beta )}}k_n^{\frac 4{3(1-\beta )}}t^{-\frac
4{3(1-\beta )}}\qquad \\ 
\text{Si }\beta =0\Rightarrow G\propto t^{-\frac 23},c\propto
t^{-1/3},\Lambda \propto t^{-4/3} 
\end{array}
$$
$c\propto t^{-1/3}$ ,this result also it is obtained by Barrow (see \cite{BA}%
) and Petit (\cite{P}) but not by Troiskii (\cite{TR}). The rest of the
quantities coincides with the model presented by Petit, except for energy
density, since Petit considers that the mass also should vary.%
$$
\rho \propto k_n^2t^{-2}\quad \quad \quad \rho \propto t^{-2}\quad 
$$
$$
f\propto A_\omega ^{\frac 1{(1-\beta )}}k_n^{-\frac 2{3(1-\beta )}}t^{\frac
2{3(1-\beta )}}\qquad \text{if }\beta =0\Rightarrow f\propto t^{2/3}\qquad 
$$
$$
\xi \propto k_n^2t^{-1}\quad \quad \quad \xi \propto t^{-1} 
$$
this model with $\ \beta =0$ is very similar to a FRW with matter
predominance.

\section{\bf Conclusions.}

We have solved through D.A. a flat model i.e. the sectional curvature of the
3-space is zero, homogeneous and isotropic i.e. we admitted symmetries type
FRW. The energy-momentum tensor is described by a fluid with bulk viscosity
in the one which furthermore we envisage mechanisms so much of creation of
matter as of entropy and in the one which the classics ''constants'' $G,c$
and $\ \Lambda $ are considered as variable.

The envisaged cases here show the following behavior for such ''constants''%
$$
G\propto t^{-1}\quad c\propto t^{-1/2}\quad \Lambda \propto t^{-1} 
$$
for a model with $\left( \gamma =1/2,\ \omega =1/3\ \text{ y }\beta
=0\right) $ and%
$$
G\propto t^{-2/3}\quad c\propto t^{-1/3}\quad \Lambda \propto t^{-4/3} 
$$
for a model with $\left( \gamma =1/2,\ \omega =0\ \text{ y }\beta =0\right) $
an equal behavior for the rest of the quantitites that are observed it in
the classic FRW.

Several problems have emerged during the development of the article.

\begin{enumerate}
\item  Even though we have supposed the classic constants of the physics as
variable have arisen (and result us indispensable) two constants, $k_\gamma $
and $A_{\omega ,\beta }$ that have clear physical meaning and without those
which we cannot arrive to solve our model. Will be these two characteristic
'' constants '' of our model also universal?

\item  For the calculation of the thermodynamic quantities as the
temperature and the entropy we have used the ''constant'' $k_B$ that we have
supposed constant and the ''constant'' of radiation $a$ that also we have
supposed constant, but if $a$ is constant (observe that $a\propto \left(
k_B^4/c^3\hbar ^3\right) $) then the only one possibility that we have is to
make $\hbar \propto c^{-1}$ that is equals to say that $\hbar \propto
t^{1/2} $ or $\hbar \propto t^{1/3}$ depending on the model.

\item  If we abandon the characteristic value of $\ \gamma =1/2$ we prove
without difficulty that for $\ \gamma >1/2$ the ''constant'' $G$ varies in a
proportional way to the time instead of inversely proportional to the time
as have obtained here. But we would need of some evidence or physical
rigorous reasoning to take similar values of $\ \gamma $. The only one
possibility to obtain $G$ and $c$ constant with $\ \gamma =1/2$ is imposing
a physically unrealistic condition $\ \omega =-1/3$.
\end{enumerate}

\end{document}